\documentclass[preprint2]{aastex}
\usepackage{graphicx}
\usepackage{epstopdf}

\shorttitle{Energetics of LMC Supernova Remnants}
\shortauthors{Leahy}

\begin{document}


\title{Energetics and Birth Rates of Supernova Remnants in the Large Magellanic Cloud}


\author{D.A. Leahy}
\affil{Department of Physics $\&$ Astronomy, University of Calgary, Calgary,
Alberta T2N 1N4, Canada}



\begin{abstract}
{Published X-ray emission properties for a sample of 50 supernova remnants (SNRs) in the LMC are used
as input for SNR evolution modelling calculations. 
The forward shock emission is modelled to obtain the initial explosion energy, age, and circumstellar medium density for each SNR in the sample.
The resulting age distribution yields a SNR birthrate of 1/(500 yr) for the LMC. 
The explosion energy distribution is well fit by a log-normal distribution, with
most-probable explosion energy of $0.5\times10^{51}$erg, with a 1-$\sigma$ dispersion by factor 3 in energy.
The circumstellar medium density distribution is broader than the explosion 
energy distribution, with most-probable density of $\sim$0.1 cm$^{-3}$. 
The shape of the density distribution can be fit with a log-normal distribution, with
incompleteness at high density caused by shorter evolution times of SNRs.
}

\end{abstract}





\section{Introduction}

The study of supernova remnants (SNRs) is of great interest in astrophysics. 
SNRs can provide valuable information relevant to their stellar progenitors and the associated explosion events that end their stellar lives. 
The hot shocked gas in the interior of a SNR is observed in X-rays, with 
a temperature $\sim$1 keV. 
This temperature is determined by the physics of shocks and the evolutionary history of the interior of the supernova remnant. 
Models, based on hydrodynamic simulations, can be constructed which represent a reasonable approximation of the evolution of a supernova remnant (e.g.  \cite{1999truelove}, \cite{1988CMB}).
Such models, together with the observed X-ray emission, can be used to deduce the 
explosion energy and the age of the SNR and the density of its surroundings. 
A study of a large number of SNRs can yield insights about the physical processes related to SNR, such as SNR evolution,
SN explosion energies and properties of the environment in which SN explode.

Studies of populations of SNRs have previously been carried out. Here a few recent studies are pointed out.
The sizes of SNRs in the LMC and SMC is known to follow a linear cumulative distribution
(\cite{2010Band} and references therein). 
\cite{2010Bad} showed this to be most likely caused by SNRs exploding in an ISM which has a broad density distribution. 
A separate study by \cite{2010Band} had a similar conclusion. 
\cite{2010Band} analysed the radio surface brightness of SNRs in relation to both SNR diameter and local ISM
density. They assumed power-law dependencies of the various quantities, including shock radius vs. time, radio surface
brightness and probability distribution of densities. By fitting the power-law models to the
observed properties of 8 Galactic SNRs, 24 LMC SNRs and 3 SMC SNRs, they were able
to put constraints on the radio emission process and show that the "constant-efficiencies" model
for radio emission did not fit the data.
\cite{2014Asvarov} studied the size distribution of SNRs in M33. SNRs were chosen based
on their X-ray hardness ratio to get a sample as free as possible from selection effects.
Monte Carlo methods were used to generate model sets of size distributions to compare to the observed
distribution. The main conclusions were that the warm ISM (with density in the range $0.1-1$ cm$^{-3}$
has a filling factor of $\sim90\%$, and the birthrate of SNRs in M33 is $\simeq$7 per 1000 yr.

An important in input required to carry out SNR modelling is the shock radius. 
This, in turn, requires knowledge of the SNR's distance. 
For this reason, this work considers SNRs located in the Large Magellanic Cloud (LMC). The distance to the LMC is well determined at 
$49.97 \pm 0.19 \rm{(statistical)} \pm 1.11$ (systematic) kpc \citep{2013Pietrz}.
LMC SNRs also have been observed extensively in X-rays by XMM-Newton \citep{2016Maggi}. 
Among their important results were: identification of SN type (core collapse or type Ia) for number of SNRs; and
measurement of LMC element abundances from SNRs dominated by emission from shocked interstellar medium.

The current paper further analyses the sample of LMC SNRs from \citet{2016Maggi} by
applying the models from \citet{1999truelove} with some additional extensions and calculations. 
Section 2 describes the data and the models that are used. 
Section 3 gives the results for the derived explosion energies, ages and circumstellar medium densities.
Section 4 summarizes the results. 

\section{LMC SNR Data and SNR Evolution Models}

\citet{2016Maggi} present the results of extensive XMM-Newton observations of LMC SNRs. 
They present X-ray images and spectral analysis results for 51 SNRs. 
From this list we analyse all but SN1987A, for which our models do not apply,
leaving a sample of 50 SNRs.
The primary data inputs for our analysis come from their Tables C1, E1 and E2.
In particular, the quantities of interest here are the outer shock radius, 
the temperature and the emission measure. 
The XMM-Newton temperatures and emission measures are generally the best available
values for these LMC SNRs.
In several cases Chandra images are available with higher spatial resolution. 
For those cases we use the shock radius from the Chandra images.

For modelling this sample, some simplifying assumptions are made. 
Spherically symmetric SNR evolution is assumed, which allows use of
analytic approximations to SNR evolution models. 
This assumption could be relaxed if carrying out multi-dimensional hydrodynamic models, but that is beyond the scope of the current work.
As discussed in \citet{1999truelove}, prior to the onset of radiative cooling,
spherically symmetric supernova remnant evolution can be described by unified
solutions. 
These are more complex than self-similar solutions, but are analytic in nature.
They describe the SNR evolution (forward and reverse shock evolution) throughout
the ejecta-dominated and Sedov-Taylor phases as well as the transition between the two phases. 
For the transition from Sedov-Taylor phase to the radiative phase the models of \citet{1988CMB} are used. 
For the circumstellar medium, a uniform density is assumed. 
The ejecta density profiles are taken as power-law in radius with index n
Truelove and McKee (1999) present models for n=0 up to n=14 ($n\ne5$). 
 $n=7$ is expected for a Type Ia explosion  \citep{1969Colgate} and 
 $n>5$ for core collapse supernovae \citep{1994ChevFr}.

The model emission measures for material heated by the forward shock, EM, 
are calculated using the interior SNR structure (density, temperature and pressure profiles): $EM= \int n_e n_H dV$.
Dimensionless EM is defined by $dEM=EM/(R_s^3 n_{e,sh} n_{H,sh})$ with post-shock electron density $n_{e,sh}=4n_{e}$ and post-shock H density $n_{H,sh}=4n_{H}$. 
$n_{H}$ and $n_{e}$ are the preshock (ISM) values, with $n_{e}$ calculated as
if the gas were ionized to the same state as the postshock gas. 
$\mu_H$ and $\mu_e$ are the mean molecular weights (or mean masses) per H atom 
and per electron of the postshock gas. 
From the definition of mean molecular weights in terms of mass density $\rho$, 
$\rho = \mu_H m_H n_{H} = \mu_e m_H n_{e}$ one obtains 
$n_{e}=\frac{\mu_H}{\mu_e} n_{H}$.
With this definition, $dEM$ is independent of $R_s$ and $n_H$, and depends only on the shape of the interior density distribution. 
$dEM$ is a constant for the self-similar phases of the evolution during which the shape of the density does not change. 
For the Sedov-Taylor phase, the interior structure and $dEM$ for the self-similar solution 
is calculated using the equations given in \citet{1991WL}
with the cloud evaporation source term set to zero.
For the ejecta dominated phase,  $dEM$ is calculated using the self-similar interior 
solutions given in \citet{1982Chevalier} for n=7 and n=12 cases.
Only the part of the self-similar solution exterior to the contact discontinuity is used 
to obtain $dEM$ of the forward-shocked material.
For the transition phase from ejecta self-similar phase to the Sedov-Taylor self-similar phase,  $dEM$ is  calculated for n=7 and 12 cases as follows. 
The ejecta self-similar phase ends at time $t_{core}$ and the Sedov-Taylor self-similar phase starts at time $t_{ST}$.
$t_{core}$ and $t_{ST}$ are defined and values for different n are given in \citet{1999truelove}.
Between $t_{core}$ and  $t_{ST}$, $dEM$  is interpolated between the ejecta phase value 
for n=7 or n=12 and the Sedov-Taylor value.
Models have been tested for different n on a few SNRs in our sample and it was
found that results differ by much less (typically less than a few parts in 1000)
than the differences caused by the uncertainties in the input parameters.
Thus the presented models are given for the n=7 case.
Tests were made using models with different values of ejected mass between 0.5 and 10 $M_{\odot}$. 
Again, the results differed by much less than differences caused by the uncertainties in the input parameters.
Thus the presented models are given for ejected mass of 1.4$M_{\odot}$. 

The temperature of the X-ray spectrum of the outer shock component of 
a SNR is the emission-weighted temperature. 
In the models considered, the temperature increases as radius decreases inside the shock front. 
Thus the emission-weighted temperature is higher than the temperature at the shock front. 
At the shock, electrons and ions are heated to different temperatures. 
The initial heating per particle for a strong shock, with adiabatic index
$\gamma=5/3$ is $\frac{3}{16} \mu m_H V_s^2$. 
Here $\mu$ is the mean molecular weight for the plasma 
$1/\mu = 1/\mu_{ion} + 1/\mu{e}$, with $\mu_{ion}$ and $\mu_{e}$, the mean 
molecular weights for the ions and for the electrons. 
Mean molecular weights corresponding to LMC abundances are used.
These affect the relation between the mass density, number density,
pressure and temperature. Thus abundances affect the shock jump conditions, SNR 
evolution, and emission measures. Details on the SNR models, including effects of abundances, are given in \cite{2017LW}. 

As the plasma ages the electrons and ions slowly equilibrate in temperature.
\cite{2013Ghav} gives a detailed discussion of electron ion equilibration.
Collisionless shocks are complex, governed by interactions in the plasma
which depend on collective processes, and electron heating is not yet well understood.
For our main calculations, the calculation of electron heating given in \cite{1982CA} 
(hereafter CA82) is followed, which uses the results of \cite{1978I}.
For simplicity, the standard composition $n_{He}=0.1 n_H$ used in CA82,
is for the calculation of electron-ion equilibration.
Observations of electron to proton temperature ratios in SNRs are relatively recent.
Thus an alternate calculation of electron heating is done here using the 
phenomenological $1/V_s^2$ model (equation 5 in \cite{2013Ghav}). 
The minimum $T_{e}/T_{ion}$ is set to 0.1 to agree with the observations,
which typically have large (factor of 2-5) errors (Fig. 2 in \cite{2013Ghav}).
The results from the two methods (CA82 and $1/V_s^2$ models) are compared. 

For the CA82 calculation, electron-ion temperature equilibration is due to
Coulomb collisions. The equilibration timescale is:
$t_{eq}=5000 E_{51}^{3/14}n_0^{-4/7}$yr, with $E_{51}$ the explosion energy in units
of $10^{51}$erg and $n_0$ the density in cm$^{-3}$.
The electron to ion temperature ratio $\alpha=T_{e}/T_{ion}$ is given to a 
good approximation by $\alpha=1-0.97exp[-(5f/3)^{0.4}(1+0.3(5f/3)^{0.6}]$. 
Here $f=\frac{ln(\Lambda)}{81}\frac{4n_0}{T_s^{3/2}}(t-t_0)$ with the Coulomb logarithm
given by $ln(\Lambda)=ln(1.2\times10^5 T_s^{1/2}T_e(4n_0)^{-1/2})$. $t_0$ is the time
a parcel of gas was shocked and the post-shock density is $4n_0$.
The factor of 0.97 in the expression for $\alpha$ has been included to give reasonable 
agreement with the measured electron to proton temperature ratios for young SNRs 
(Fig.2 in \cite{2013Ghav}).
For typical SNR parameters (explosion energy $10^{51}$ erg, ejected mass of 1.4$M_{\odot}$, density of 1 cm$^{-3}$) 
our calculations gives $\alpha \simeq 0.055$ for age of 100 yr,
increasing rapidly to 0.31 at age of 1000 yr, then more slowly, reaching 0.97 at 5000 yr.  
To get realistic X-ray (i.e. electron) temperatures from the models, the inclusion of progressive equilibration of electron and ion temperatures is essential.

The ionization states of the ions are generally out of equilibrium with the electron temperature for many SNRs.
However, this is already taken into account in modelling the X-ray spectra, so that
the electron temperatures derived from the X-ray spectrum are corrected for this effect.

\section{Results and Discussion}

For each SNR in the LMC with observed radius, R, emission measure, EM, and X-ray temperature, kT, 
an initial explosion energy, $E_0$, age, and circumstellar medium density, $n_0$, were used to calculate a model. 
The process was iterated until convergence of the output R, EM and kT to the observed values.
The procedure is similar to that described in \citet{2016LeahyRana}.
The resulting $E_0$, age and $n_0$ for each of the 50 SNRs is listed in Table 1. 
The input R is also given. 
The input EM and kT and their uncertainties are identical to those listed in 
 \citep{2016Maggi} so are not repeated here.
The results from model fitting using the alternate prescription for electron heating 
show the following differences. Derived density, $n_0$, is identical. 
Age on average was smaller by a factor of 0.88 (standard deviation 0.19) 
and energy larger by a factor of 1.22 (standard deviation 0.52).
For SNRs with X-ray temperature $<0.27$keV (13 SNRs) there is no
significant difference ($<1\%$ and typically $<0.1\%$) in age or explosion energy. 
For X-ray temperature $>0.27$keV (37 SNRs), the age is usually smaller,
and explosion energy is usually larger.
The mean log(age) is 4.12 ($1.3\times10^4$yr) for the CA82 $T_e/T_{ion}$ method and 
4.05 ($1.1\times10^4$yr) for the $1/V_s^2$ $T_e/T_{ion}$ method.
The mean log($E51$) is -0.293 ($5.1\times10^{50}$erg) for the CA82 $T_e/T_{ion}$ method and 
-0.144 ($7.2\times10^{50}$erg) for the $1/V_s^2$ $T_e/T_{ion}$ method.
  
To obtain uncertainties in $E_0$, $n_0$, and age, the models were rerun with 
EM and kT set to the 4 different combinations of upper and lower limits. 
Then the extreme values of $E_0$, age and $n_0$ from this set of fits were used as the upper and lower limits.  
 Generally asymmetric errors on the parameters were obtained, as shown in Table 1 here.
The errors in Table one are derived from the errors of the input parameters. An
additional systematic error for $E_0$ and age exists because of the uncertainty in 
 $T_e/T_{ion}$ calculation. A measure of that uncertainty is the standard deviation of
difference in results from the two $T_e/T_{ion}$ calculation methods, 
which is $19\%$ for age and $52\%$ for $E_0$.

There are wide ranges for the deduced values of $E_0$, age and $n_0$. 
The supernova (SN) explosion energies range from 0.037$\times10^{51}$ erg (J0453-6829) to 11$\times10^{51}$ erg (J0449-6920), i.e by a factor of 300. 
Such a wide range should not be surprising: the range in observed SN energies is 
similarly wide (for Type II SN energies see \citet{2016Rubin}).  
The deduced ages range from 1300 yr (J0537-6910) to 60,000 yr (J0517-6759). 
This is in rough agreement with what is expected because SNRs are expected to become too
faint to observe after several $10^4$ yr. 
The deduced circumstellar densities range from 0.0007 $cm^{-3}$ (J0505-6753) to 5.5 $cm^{-3}$ (J0453-6655).
This spans 4 orders of magnitude and is similar to what is expected for Milky Way SNRs exploding
in different environments, ranging from the hot ionized medium, with density $\simeq 0.001 cm^{-3}$,
to the diffuse warm and cool HI, with density $\simeq 0.1-1 cm^{-3}$, 
to the molecular medium with density $\geq 1 cm^{-3}$ \citep{2005Cox}.

Plots of the various SNR parameters were made to check that the models gave reasonable results.
Fig. 1 shows  $n_0$ vs. age. The SNRs are scattered across the plot except for a
clear deficit in the upper-right corner (high density and high age). This absence
is expected because the onset of radiative losses occurs earlier for SNR in higher density
ISM. The expected transition time to the PDS phase is 
$t_{PDS}=13300(E51)^{3/14}n_0^{-4/7}\zeta_m^{-5/14}$ (CMB88), with $\zeta_m$ the 
metallicity correction to the cooling function. This is plotted in Fig. 1 as the dashed 
and dotted lines for explosion energies of $10^{50}$ and $10^{51}$ erg. SNRs become
much fainter as they approach the PDS phase, so the upper limits on the derived ages are 
consistent with expectations.
A plot of explosion energy vs. age shows no correlation. 
However there is an empty region in the plot at large age and low explosion energy,
which is likely because old SNRs with low explosion energy are too faint to be in the sample.
The plot of radius vs. age is similar to the plot of radius vs. number because the birthrate
of SNRs in the LMC is consistent with a constant value (see below). 
The reason for the nearly linear trend was discussed in detail by \cite{2010Bad}: 
it can be well explained by the spread of ISM densities in which SNR explode.

Fig. 2 shows the distribution of SN explosion energies, $E_0$, derived here for the LMC SNRs. 
$E_0$ for each SNR is plotted as a function of cumulative number of SNRs with energy less than
or equal to $E_0$, which is equal to the rank number by energy. 
The energies are clearly not uniformly distributed, otherwise the slope would be
a straight line. 
Other simple distribution functions, such as Gaussian in energy, were tested and clearly do not
fit the LMC sample values of $E_0$.
Instead the distribution is reminiscent of a log-normal probability distribution:
\begin{equation}
\frac{dN}{dE} = \frac{N_T}{\sqrt{2\pi \sigma_{logE} ^2}} e^{-\frac{[log(E)-log(E_{av})]^2}{2 \sigma_{logE}^2}}
\end{equation}
where $N_T$ is the total number in the sample, $E_{av}$ is peak in number vs. energy and 
$\sigma_{logE}$ the dispersion in log(E). 
The solid line in Fig. 1 is the best fit cumulative distribution, with parameters  
$E_{av}= 0.48\times10^{51}$erg and $\sigma_{logE}$=0.47. The latter corresponds
to a 1-$\sigma$ dispersion of a factor 2.94 in energy.
The process was repeated for the explosion energies derived using the 
 the $1/V_s^2$ $T_e/T_{ion}$ method. The plot looks nearly the same as the one in Fig.2.
 Those energies are also well-fit by a log-normal distribution, but with parameters
 $E_{av}= 0.54\times10^{51}$erg and $\sigma_{logE}$=0.50. The latter corresponds
to a 1-$\sigma$ dispersion of a factor 3.16 in energy.

The distribution of SN ages for the LMC SNRs is shown in Fig. 3. 
Age is plotted for each SNR vs. cumulative number of SNRs with age less than
or equal to that of the given SNR. 
In this case, for ages $\le$20,000 yr, the distribution is uniformly distributed in age, and
is well fit by a straight line.
The slope of the line is the birthrate of SNRs in our sample, which corresponds to observable X-ray SNRs in the LMC.
The derived birthrate is 1/(503yr), where the fit is done only for SNRs with age $\le$20,000 yr. 
The fits were repeated for the ages derived using the 
 the $1/V_s^2$ $T_e/T_{ion}$ method. The plot looks nearly the same as the one in Fig.3.
 The derived birthrate is 1/(507yr), where again the fit is done only for SNRs with age $\le$20,000 yr.

The excess of ages for rank $>$40 can be attributed to incompleteness. 
I.e., because the SNRs with age$>$20,000 yr are becoming faint, the X-ray surveys are seeing
only a fraction of the population. 
If all of the age$>$20,000 yr SNRs (say up to $6\times10^4$yr) were observed, then the rank
numbers would be larger, flattening the slope. 
The incompleteness is estimated by fitting a slope to the set with age rank between 40 and 50. 
This gives a birthrate of 1/(4200yr), suggesting an incompleteness of a factor of $\simeq$8
for SNRs with age between 20,000 and 60,000 yr compared to those with age $\le$20,000 yr.
Including SN1987A in the sample has the following effect: 
the birthrate remains at 1/(503 yr) but y-intercept of the line becomes 1060 yr instead of
1560 yr. This suggests that $\sim$2 other young SNRs are missing in the sample, possibly because confusion because of their small size.
If the assumption is made that every observable SN produces an observable SNR (for age $\le$20,000 yr),
then the SN rate for the LMC is 1/(500 yr), which is about 1/10 that of the SN rate for the Milky
Way. This is roughly consistent with the stellar luminosity of the LMC which is about 1/10 that
of the Milky Way. The uncertainties in the SN rate of the Milky Way and the stellar luminosities of
the LMC and Milky Way are large enough that a more detailed comparison is not possible.

As a separate test of the derived birthrate, the derived ages were replaced with available estimates from the literature, as summarized in \citet{2016Maggi}. 
Previous estimates were available for 25 LMC SNRs.
It was not assessed for each case whether the previous age estimates are more reliable or less reliable than the currently derived ages, 
in fact many of them are consistent with the same value. A detailed comparison for individual SNRs will be done in future work.
The resulting cumulative age distribution looks much the same as that shown in Fig. 2,
but the fitted slope is lower, corresponding to a birth-rate of 1/(607 yr). 
The difference is taken as a measure of the uncertainty in the LMC SNR birthrate, i.e.
$\sim$1/(500 yr) to $\sim$1/(600 yr).
Future assessment of the methods of obtaining ages should reduce this uncertainty. 

Fig. 4 shows the distribution of SNR circumstellar densities, $n_0$. 
The densities are clearly not uniformly distributed. 
Here the $n_0$ distribution is fit with a log-normal probability function:
\begin{equation}
\frac{dN}{dn_0} = \frac{N_T}{\sqrt{2\pi \sigma_{log(n_0)}^2}} e^{-\frac{[log(n_0)-log(n_{0,av})]^2}{2 \sigma_{log(n_0)}^2}}
\end{equation}
where $n_{0,av}$ is peak in number vs. density and 
$\sigma_{log(n_0)}$ the dispersion in log($n_0$). 
The fit with a single log-normal distribution is poor but separate fits to the lower and
upper sets of densities is good.
The solid blue line is the best fit to the 30 lowest densities, with parameters  
$n_{0,av}= 0.079 cm^{-3}$ and $\sigma_{log(n_0}$=0.51. 
The dashed red line is the best fit to the 30 highest densities, with parameters  
$n_{0,av}= 0.091 cm^{-3}$ and $\sigma_{log(n_0)}$=0.91. 
The $\sigma$'s correspond to 1-$\sigma$ dispersions of by factors of 3.2 and 8.2 in density.
Because the peak densities for both fits are so close to the same value, this
suggests that there are not 2 physically distinct distributions of density. 
One possibility density distribution is not well-described by a log-normal distribution.
This may not be surprising considering the complexity of processes that
govern the densities in the ISM (e.g. see \citet{2005Cox}).
Another possibility is that the high density end of the distribution is under-sampled.
SNRs occurring in a high density medium evolve faster and become radiative
at earlier times. 
Thus they would be under-represented in the sample compared to longer-lived
SNRs occurring in lower density environments. 
In this case, the poor fit above density of $\sim0.2~cm^{-3}$ would be explained by incompleteness.
The mean age for the 30 SNRs with lowest density is 21,000 yr, whereas the
mean age of the 20 SNRs with highest density is 12,000 yr. This supports
the suggestion that the high density end of the sample is affected by incompleteness.

The density distribution at the sites of SNRs in the LMC has been discussed previously.
\cite{2010Bad} and \cite{2010Band} study the size distributions of SNRs in the LMC, SMC and M33.
There is a nearly linear size distribution over a fairly wide range in radius (up to 60 pc).
Both studies consider SNRs which evolve in a distribution of densities and use this
to explain the linear size distribution. 
The upper cutoff in the size distribution is explained by a minimum in the density distribution (\cite{2010Bad})
Our result of an approximately log-normal distribution for density is consistent with
the linear size distribution of LMC SNRs, because it was derived from nearly the same sample
of LMC SNRS as used by \cite{2010Bad}. 
The density distribution derived here (Fig.3) has been fitted by a $n_0^{-1}$ distribution (the green line in Fig.3),
however the log-normal distribution gives a significantly better fit.
The reasons that the density distribution here is different the one in \cite{2010Bad} are: 
they assumed a power-law density distribution; individual SNRs were modelled here with 
explosion energy, age and density as parameters for each SNR.   

For the current work, the X-ray emission of the forward shock was analysed.  
It was noted that the results are very insensitive to the amount of ejected mass, $M_{ej}$, 
and to the index n for the power-law density distribution of the ejecta. 
However, the behaviour of the reverse shock is sensitive to both parameters. 
The SNR evolution model is currently being improved to be able to model the
reverse shocked ejecta for values of n between 0 and 14. 
When the reverse shock model is completed, the X-ray emission from the reverse
shock, which is seen in a number of the LMC SNRs, can be modelled.
That modelling can yield $M_{ej}$ and n values for a sample of SNRs and will
enable us to learn about the properties of the SN progenitors.

\section{Summary}

X-ray emission properties from the forward shock for a set of 50 SNRs in the LMC has 
been determined recently by \citet{2016Maggi}. 
Here, SNR evolution and interior structure calculations have been carried out in
order to match the observed radius, emission measure and electron temperature for
each of the 50 SNRs. 
The results for explosion energy, age and circumstellar medium density, and their 
uncertainties, are given in Table 1. 
$T_e/T_{ion}$ was calculated using the \cite{1982CA} prescription 
and an additional set of models was calculated using the $1/V_s^2$ prescription (\cite{2013Ghav}). The latter gave identical densities, and different ages and explosion
energies. However, the analysis of the distributions of ages and explosion energies gave the
same results using both methods. 

This is the first time that the energy distribution of SNRs, the birthrate of SNRs and the density distribution for SNRs have been measured for any large sample.
The distribution of parameters is summarized in Figs. 2, 3 and 4.
For  explosion energy and density, the distributions were fit by log-normal distributions.
A most-probable explosion energy of $0.5\times10^{51}$erg is found, with a 1-$\sigma$ dispersion by factor 3 in energy.
For density, two log-normal fits are better than one, which may be caused by complexity in 
the distribution of interstellar medium in the LMC or by incompleteness in the sample. 
In either case, the mean density is $\sim0.1 cm^{-3}$ with a 1-$\sigma$ dispersion by factor 
$\sim$3-8 in density.
For age, incompleteness is clearly a factor for the older part of the sample (age $>$20,000 yr).
For age $\le$20,000 yr, the ages are well fit by a constant birthrate of 1/(500 yr).
 
It would be highly desirable to carry out a similar study for Galactic SNRs. 
One of the main factors limiting such a study is lack of reliable distances to SNRs in
the Galaxy. However, once a sample of Galactic SNRs with distances and X-ray observations
is obtained, similar modelling can be carried out.
Then the properties of Galactic and LMC SNRs can be compared.

\acknowledgments

D.L. acknowledges the assistance of undergraduate student J.E. Williams, who 
compiled the LMC data used as input for this study, and  also verified 
some of the model calculations. 
This work was supported by a grant from the Natural Sciences and Engineering Research Council of Canada. 

\clearpage



\begin{figure}
\epsscale{1.}
\plotone{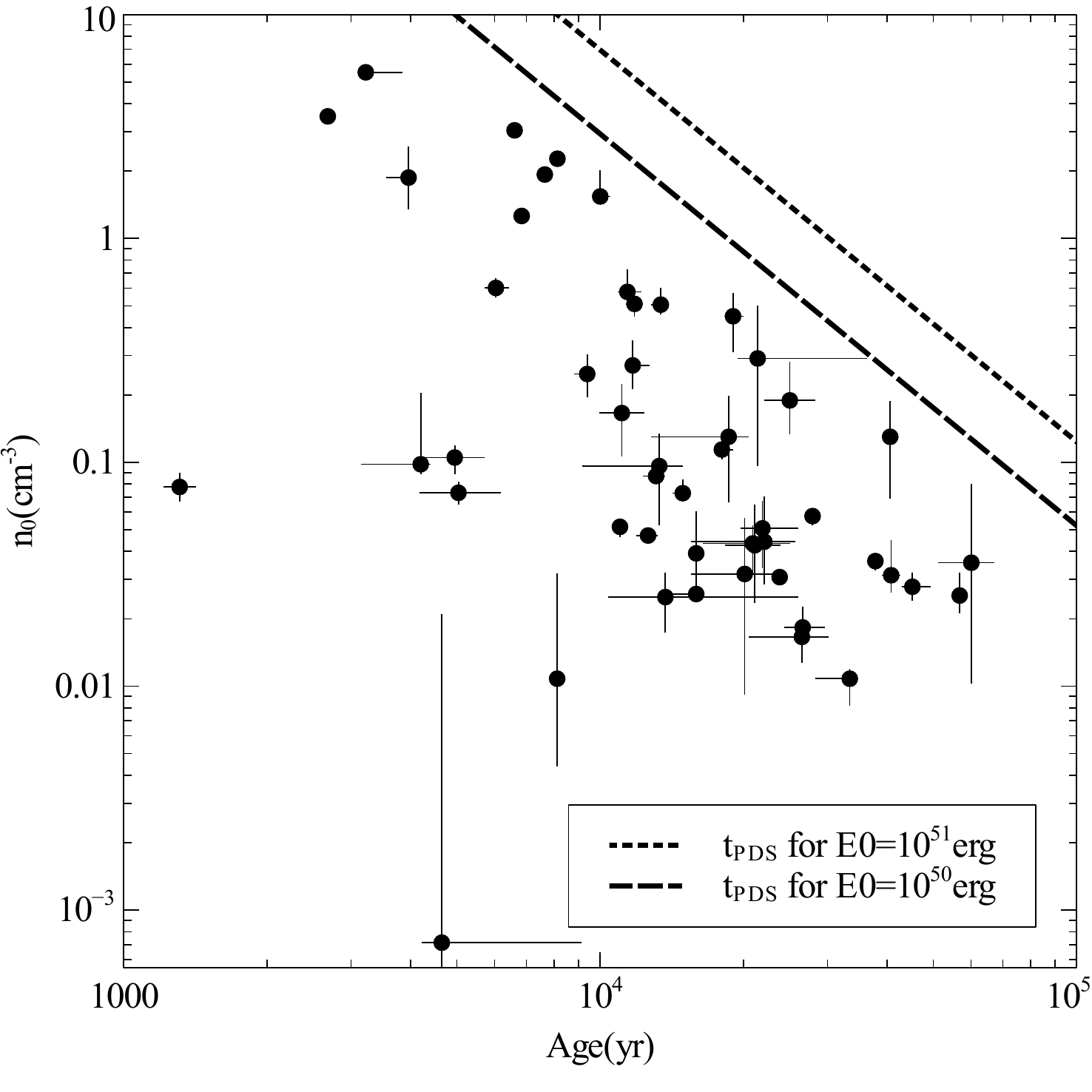}
\caption{The derived densities vs. ages for the LMC SNRs (circles with error bars). The dashed
and dotted lines show the theoretical transition time to the pressure-driven snowplow (PDS)
phase, caused by radiative losses, for explosion energies of $10^{50}$ and $10^{51}$ erg.\label{fig0}}
\end{figure}
\clearpage

\begin{figure}
\epsscale{1.}
\plotone{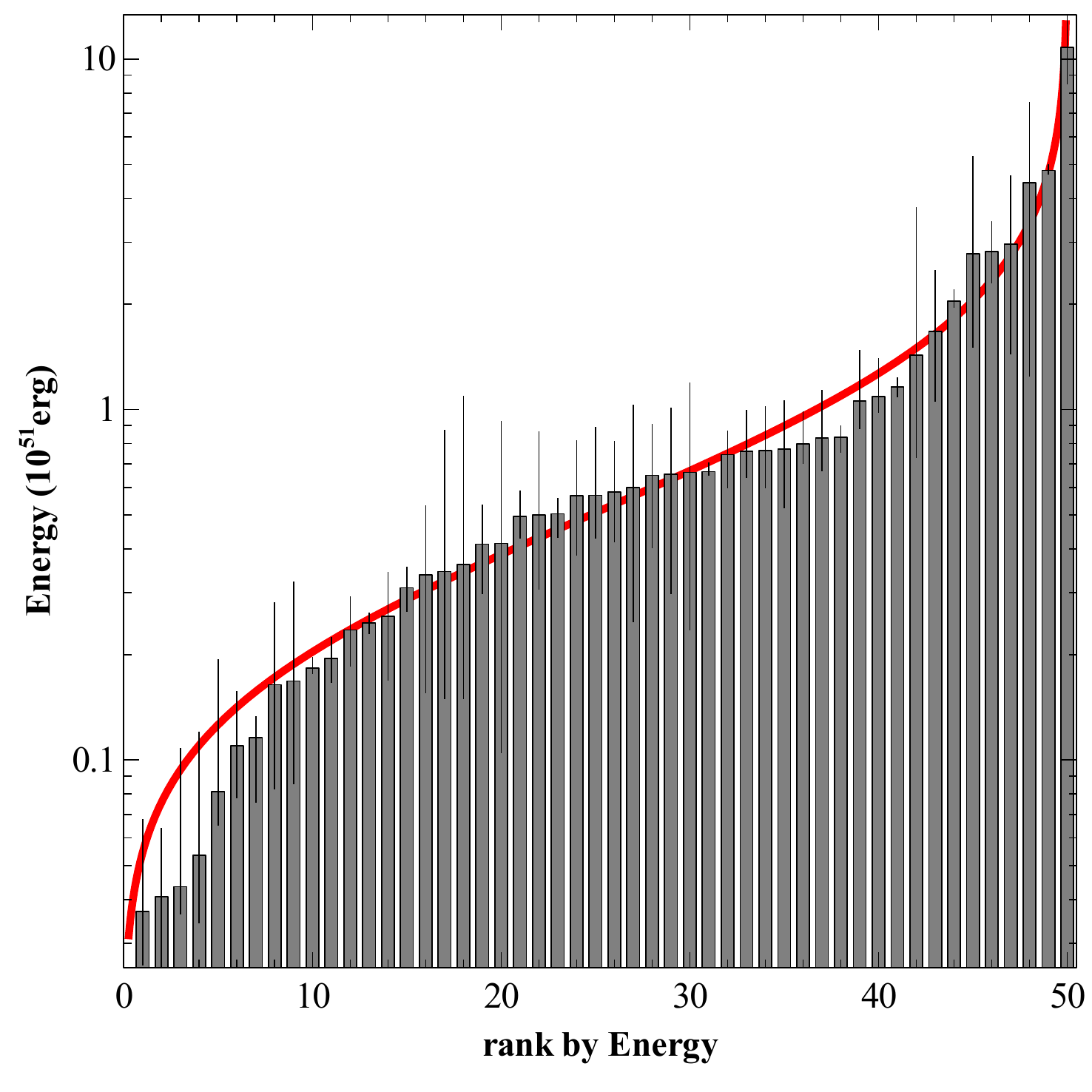}
\caption{The cumulative distribution of explosion energies (histogram with error bars) for the LMC supernova remnant sample. The horizontal axis is the cumulative number of SNRs with energy less than or equal to the energy of a given SNR, the vertical axis is the energy of that SNR. The solid curve is a fit for the probability distribution expressed as a log-normal distribution (parameters given in the text).\label{fig1}}
\end{figure}
\clearpage

\begin{figure}
\epsscale{1.}
\plotone{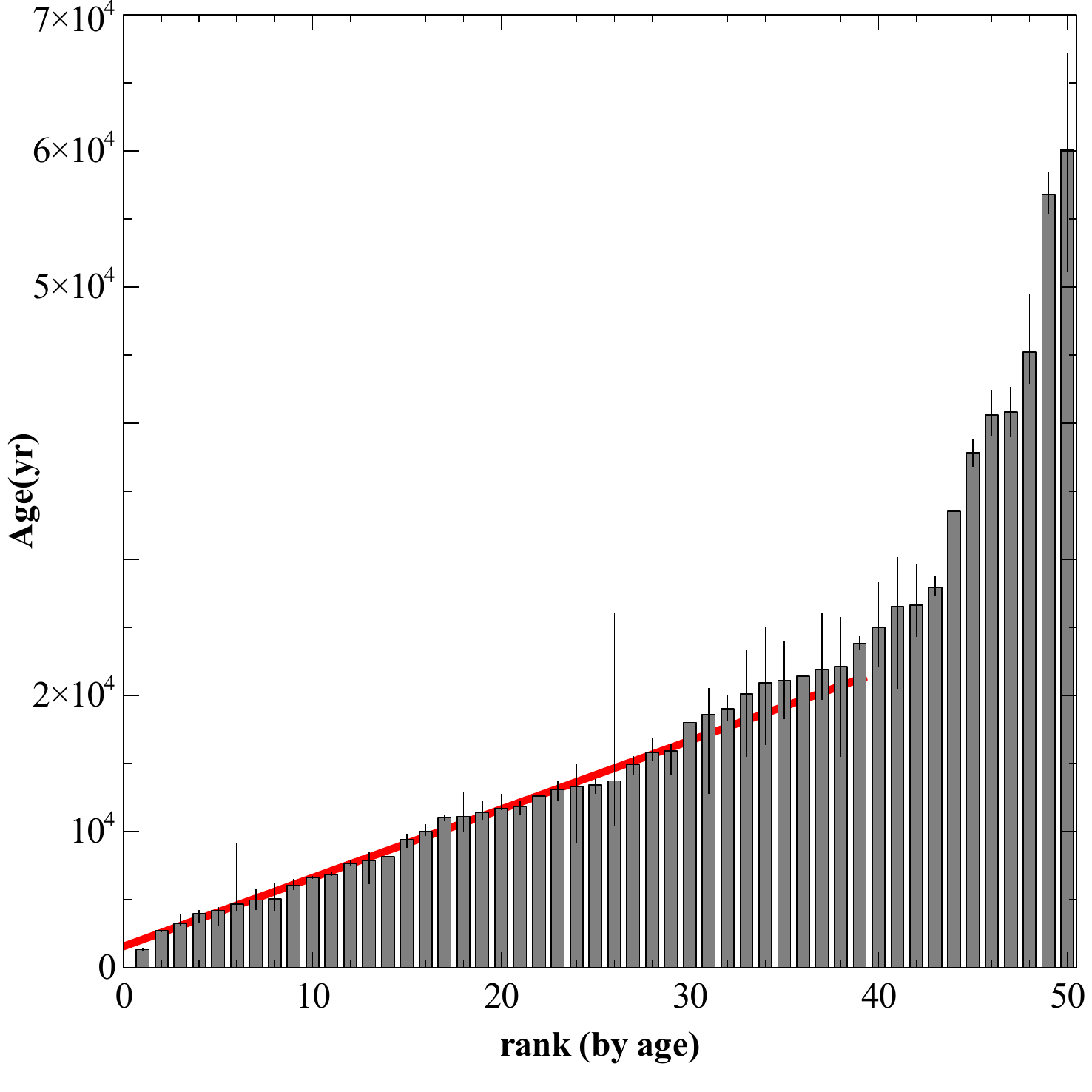}
\caption{The cumulative distribution of ages (histogram with error bars) for the LMC supernova remnant sample. The solid curve is a linear fit equivalent to a constant birthrate of 1 per 503 yr. The fit is done for the 40 youngest SNRs in the sample.\label{fig2}}
\end{figure}
\clearpage

\begin{figure}
\epsscale{1.}
\plotone{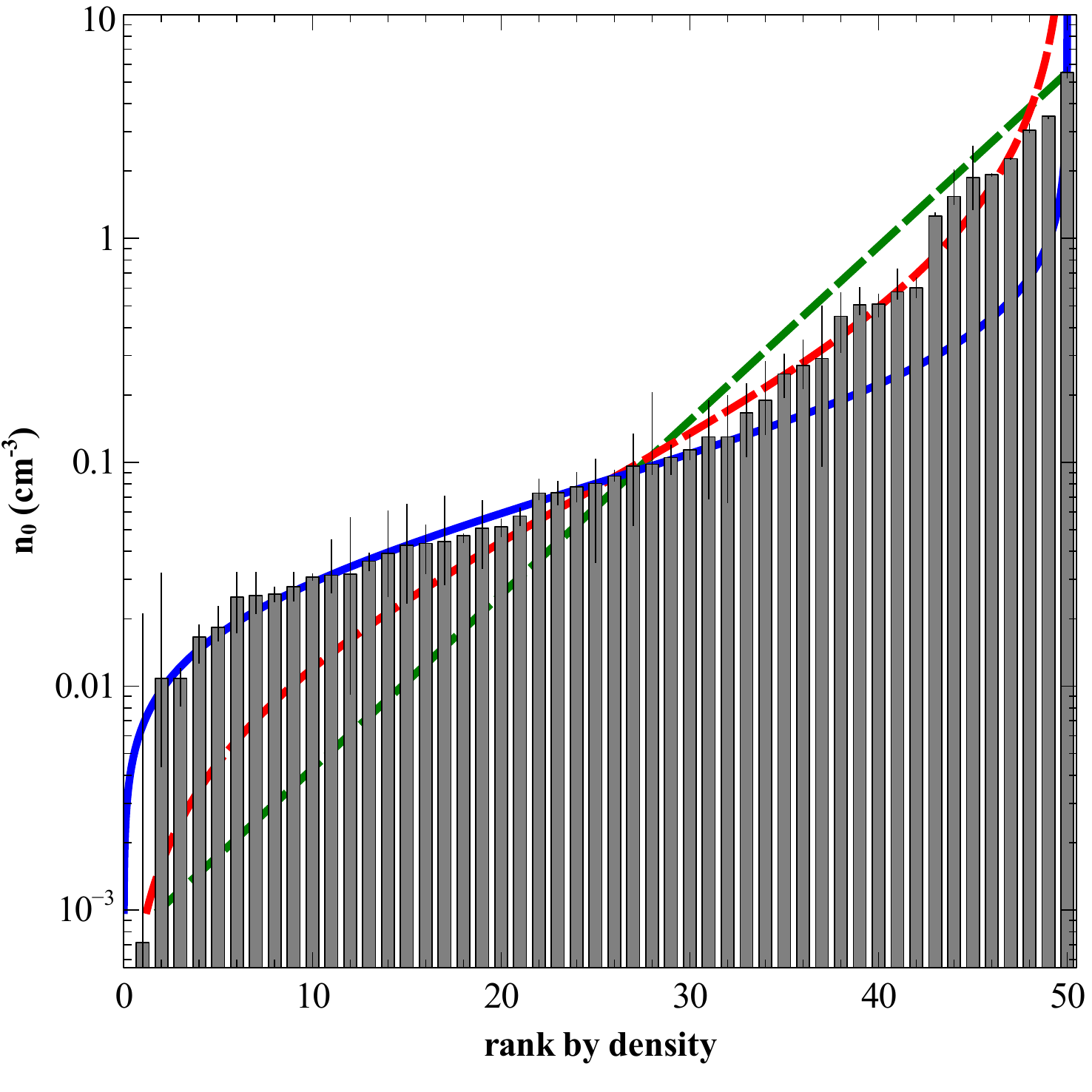}
\caption{The cumulative distribution of circumstellar densities (histogram with error bars) for the LMC supernova remnant sample. The red and blue curves are fits to the probability distribution by a log-normal distribution. The solid blue curve is the fit to the low density part of the sample (lowest 30 densities) and the dashed red curve is a fit to the high density (highest 30 densities) part of the sample. A significantly worse fit to the density is by a $n_0^{-1}$ probability
distribution, shown by the green line.\label{fig3}}
\end{figure}
\clearpage







\clearpage

\begin{deluxetable}{crrrrrrrrrrr}
\tabletypesize{\scriptsize}
\tablecaption{Models for LMC SNRs$^{a}$}
\tablewidth{0pt}
\tablehead{
\colhead{MCSNR} & \colhead{R(pc)} & \colhead{$E_0$($10^{51}$erg)} & \colhead{$E_0$ error} 
& \colhead{age(yr)} & \colhead{age error} &  \colhead{$n_0$(cm$^{-3}$)} &
 \colhead{$n_0$ error}
}
\startdata
J0449-6920 & 19.6 & 0.599 & 0.149/-0.351 & 21400 & 14900/-2000 & 0.291 & 0.209/-0.195 \\
J0450-7050 & 41.2 & 2.96 & 1.33/-1.25 & 40600 & 1800/-1500 & 0.13 & 0.058/-0.0611 \\
J0453-6655 & 31.0 & 2.78 & 0.42/-0.28 & 25000 & 3300/-2900 & 0.189 & 0.092/-0.056 \\
J0453-6829 & 14.2 & 0.829 & 0.221/-0.112 & 11400 & 800/-500 & 0.577 & 0.153/-0.039 \\
J0454-6626 & 12.8 & 0.164 & 0.057/-0.058 & 11100 & 1300/-1160 & 0.166 & 0.058/-0.06 \\
J0505-6753 & 9.45 & 0.674 & 0.004/-0.009 & 6840 & 80/-60 & 1.28 & 0.02/-0.02 \\
J0505-6802 & 11.6 & 1.06 & 0.2/-0.02 & 10000 & 500/-290 & 1.54 & 0.48/-0.12 \\
J0506-6541 & 49.7 & 0.765 & 0.147/-0.095 & 56800 & 1600/-1400 & 0.0254 & 0.0067/-0.0043 \\
J0506-7026 & 31.8 & 1.1 & 0.17/-0.01 & 15900 & 500/-1700 & 0.0258 & 0.0018/-0.002 \\
J0508-6902 & 36.8 & 0.582 & 0.024/-0.01 & 26600 & 3000/-2300 & 0.0183 & 0.0043/-0.0024 \\
J0508-6830 & 16.7 & 0.0871 & 0.0509/-0.0058 & 4650 & 4480/-430 & .000715 & 0.0203/-0.000161 \\
J0509-6844 & 3.60 & 0.116 & 0.003/-0.0312 & 3220 & 630/-120 & 5.52 & 0.29/-0.32 \\
J0509-6731 & 3.75 & 0.037 & 0.0135/-0.01 & 3960 & 150/-400 & 1.87 & 0.7/-0.52 \\
J0511-6759 & 13.6 & 0.0535 & 0.0448/-0.0191 & 8120 & 280/-270 & 0.0108 & 0.0212/-0.00642 \\
J0512-6707 & 14.5 & 0.0407 & 0.0178/-0.0126 & 15900 & 100/-100 & 0.0392 & 0.0213/-0.0141 \\
J0513-6912 & 29.1 & 0.654 & 0.114/-0.16 & 21100 & 2800/-2800 & 0.0426 & 0.0221/-0.0191 \\
J0514-6840 & 26.7 & 0.246 & 0.002/-0.003 & 23800 & 500/-400 & 0.0307 & 0.0009/-0.0009 \\
J0517-6759 & 39.3 & 0.662 & 0.018/-0.251 & 60100 & 7000/-9000 & 0.0356 & 0.0448/-0.0253 \\
J0518-6939 & 17.9 & 0.37 & 0.039/-0.025 & 13300 & 1600/-4130 & 0.0962 & 0.0378/-0.044 \\
J0519-6902 & 4.00 & 0.183 & 0.013/-0.001 & 2680 & 20/-40 & 3.51 & 0.02/-0.08 \\
J0519-6926 & 23.0 & 0.744 & 0.032/-0.072 & 18000 & 1000/-100 & 0.114 & 0.019/-0.011 \\
J0523-6753 & 21.8 & 0.504 & 0.017/-0.04 & 12600 & 600/-700 & 0.0471 & 0.0007/-0.0033 \\
J0525-6938 & 13.4 & 4.81 & 0.13/-0.04 & 8130 & 70/-120 & 2.27 & 0.01/-0.01 \\
J0525-6559 & 18.9 & 2.82 & 0.09/-0.08 & 11800 & 400/-500 & 0.51 & 0.054/-0.062 \\
J0526-6605 & 10.2 & 1.16 & 0.04/-0.04 & 7650 & 190/-180 & 1.93 & 0.02/-0.03 \\
J0527-6912 & 24.0 & 0.195 & 0.005/-0.008 & 27900 & 800/-600 & 0.0575 & 0.0049/-0.0052 \\
J0527-6714 & 32.7 & 0.31 & 0.008/-0.01 & 37800 & 1000/-1000 & 0.0362 & 0.003/-0.0034 \\
J0527-7104 & 44.7 & 0.568 & 0.037/-0.035 & 33400 & 100/-5100 & 0.0108 & 0.0012/-0.00265 \\
J0528-6727 & 39.3 & 0.569 & 0.172/-0.049 & 40800 & 1800/-1800 & 0.0313 & 0.0136/-0.0052 \\
J0529-6653 & 17.6 & 1.67 & 0.51/-0.44 & 5040 & 1150/-870 & 0.0731 & 0.0091/-0.0084 \\
J0530-7008 & 39.4 & 0.4 & 0.013/-0.002 & 45200 & 4200/-2300 & 0.0278 & 0.0043/-0.0038 \\
J0531-7100 & 19.6 & 0.496 & 0.035/-0.019 & 13100 & 600/-800 & 0.0868 & 0.0054/-0.0044 \\
J0532-6732 & 34.5 & 1.68 & 0.23/-0.25 & 22100 & 3600/-6600 & 0.0442 & 0.0261/-0.0157 \\
J0533-7202 & 24.9 & 0.31 & 0.028/-0.031 & 21900 & 4100/-2200 & 0.0507 & 0.0168/-0.017 \\
J0534-6955 & 15.3 & 0.76 & 0.086/-0.007 & 13400 & 400/-600 & 0.506 & 0.096/-0.049 \\
J0534-7033 & 22.5 & 0.833 & 0.031/-0.045 & 11000 & 200/-200 & 0.0515 & 0.004/-0.005 \\
J0535-6602 & 9.80 & 2.04 & 0.08/-0.01 & 6610 & 60/-60 & 3.04 & 0.19/-0.06 \\
J0535-6918 & 10.2 & 0.11 & 0.012/-0.009 & 9390 & 400/-580 & 0.248 & 0.055/-0.053 \\
J0536-6735 & 23.6 & 0.797 & 0.05/-0.007 & 14900 & 600/-700 & 0.0728 & 0.0113/-0.0045 \\
J0536-7039 & 21.8 & 0.168 & 0.038/-0.027 & 20900 & 4100/-4500 & 0.0434 & 0.0091/-0.0116 \\
J0536-6913 & 6.54 & 0.0435 & 0.0301/-0.0035 & 4200 & 210/-1050 & 0.098 & 0.107/-0.0095 \\
J0537-6628 & 27.5 & 0.415 & 0.12/-0.191 & 20100 & 3200/-4600 & 0.0317 & 0.0248/-0.0225 \\
J0537-6910 & 14.5 & 10.8 & 0.1/-0.4 & 1310 & 110/-100 & 0.0776 & 0.0122/-0.011 \\
J0540-6944 & 17.5 & 0.649 & 0.093/-0.156 & 19000 & 1000/-800 & 0.449 & 0.121/-0.139 \\
J0540-6920 & 8.36 & 0.235 & 0.005/-0.008 & 6040 & 390/-330 & 0.601 & 0.062/-0.057 \\
J0541-6659 & 36.4 & 0.5 & 0.146/-0.068 & 26500 & 3600/-6000 & 0.0166 & 0.0021/-0.0039 \\
J0543-6858 & 40.0 & 4.44 & 0.95/-2.43 & 13700 & 12300/-3300 & 0.025 & 0.0072/-0.0077 \\
J0547-6943 & 19.6 & 0.362 & 0.086/-0.042 & 18600 & 1900/-5800 & 0.13 & 0.068/-0.0638 \\
J0547-6941 & 13.8 & 0.771 & 0.109/-0.122 & 4950 & 760/-660 & 0.105 & 0.014/-0.0166 \\
J0547-7025 & 13.1 & 0.257 & 0.021/-0.049 & 11700 & 1000/-100 & 0.271 & 0.079/-0.058 \\

\enddata

\tablenotetext{a}{All models have SN ejecta mass of 1.4$M_{\odot}$. 
$R$ is the outer shock radius, $E_0$ is the explosion energy, $n_0$ is the pre-shock density.}
\end{deluxetable}


\end{document}